\documentclass[letterpaper]{article}
\usepackage{aaai}
\usepackage{times}
\usepackage{helvet}
\usepackage{courier}
\usepackage{tabularx}
\usepackage{ragged2e}
\usepackage{booktabs}
\usepackage{graphicx}
\usepackage{url}
\begin{document}
%
\title{Governance at the Edge of Architecture: Regulating NeuroAI and Neuromorphic Systems}

\author{Afifah Kashif\\
University of Cambridge\\
Cambridge, UK\\
ak2687@cam.ac.uk\\
\And
Abdul Muhsin Hameed\\
University of Washington\\
Seattle, USA\\
muhsinh@uw.edu\\
\And
Asim Iqbal\\
Tibbling Technologies\\
Seattle, USA\\
asim@tibbtech.com \\
}

\maketitle
\begin{abstract}
\begin{quote}
Current AI governance frameworks, including regulatory benchmarks for accuracy, latency, and energy efficiency, are built for static, centrally trained artificial neural networks on von Neumann hardware. NeuroAI systems, embodied in neuromorphic hardware and implemented via spiking neural networks, break these assumptions. This paper examines the limitations of current AI governance frameworks for NeuroAI, arguing that assurance and audit methods must co-evolve with these architectures, aligning traditional regulatory metrics with the physics, learning dynamics, and embodied efficiency of brain-inspired computation to enable technically grounded assurance.
\end{quote}
\end{abstract}

\section{Introduction}
\label{sec:intro}
NeuroAI refers to the bidirectional research agenda linking neuroscience and artificial intelligence by using brain insights to build smarter, more efficient systems, and using AI to model and test brain function. It represents not only a technical convergence but also a conceptual one, reimagining intelligence as an embodied, energy-efficient, and continuously adaptive process. Neuromorphic computing is the hardware instantiation of this agenda, departing from the previous standard of von Neumann architecture by co-locating memory and computation and operating asynchronously and event-driven \cite{kimovski_beyond_2024}.

At the algorithmic level, this shift is embodied in spiking neural networks (SNNs), which aim to capture the temporal and event-based nature of biological signaling. SNNs communicate via discrete spikes whose information is encoded in rate and timing, often supporting local learning rules such as spike-timing-dependent plasticity \cite{ghosh2009spiking}. Artificial neural networks (ANNs), like those used in applications such as ChatGPT, are widely recognized. In contrast, ANNs use continuous activations and global error backpropagation, forming the backbone of modern deep learning \cite{zou2020hybrid}. 

Collectively, NeuroAI serves as the scientific bridge, where neuromorphic computing provides the hardware foundation, and SNNs and ANNs represent the algorithmic spectrum through which this vision is realized. Together, this bridge signals a paradigm shift in which neural computation can empower learning and data processing on future hardware. For instance, neuromorphic event-driven vision systems are now being explored for implantable or portable health monitors that detect neural or cardiovascular anomalies in real time, with power requirements low enough for continuous edge deployment \cite{DeLuca2025}. 

In recent years, governments and organizations around the world have rushed to establish frameworks for regulating AI \cite{Fjeld2020-oq,cole2024ai,Maslej2025AIIndex}. Common examples include the EU AI Act \cite{eu_ai_act}, U.S. NIST AI Risk Management Framework \cite{NIST} and China AI Safety Governance Framework \cite{china_framework}. These frameworks, among many others, range from binding legislation to voluntary guidance, often reflecting national values, economic priorities, and governance models. As NeuroAI develops, governance must also evolve, as technical progress cannot occur in isolation from considerations of safety and societal impact and principles of transparency, auditability, and trustworthiness are embedded into the design of algorithms and hardware from the outset. Without this co-development, regulatory mechanisms risk lagging behind technological realities, creating gaps that could hinder both innovation and public confidence \cite{reuel2024positionpapertechnicalresearch}.

Emerging frameworks such as NeuroBench propose a device-agnostic framework spanning task, model, and platform layers, reporting not only accuracy but also latency, power, and energy per sample with open reference code. By tying algorithmic performance to hardware efficiency, NeuroBench reframes evaluation from raw compute to a systems-level audit \cite{neurobench}. This paper extends this foundation into governance. 

This paper explores how metrics of efficiency, adaptability, and embodiment can be translated into regulatory language—a step that becomes increasingly necessary for NeuroAI to transition responsibly from lab prototypes to real-world systems. It examines how existing AI governance frameworks, designed for static, high-compute, centrally trained models, fail to capture the adaptive and event-driven behavior of neuromorphic and NeuroAI systems. The analysis demonstrates that assurance and audit methods must co-evolve with these architectures, aligning traditional regulatory metrics with the physics and learning dynamics of brain-inspired computation.

\section{Current Regulatory Frameworks}
\label{sec:curr_frameworks}
Around the world, governments and international bodies have proposed or enacted regulatory frameworks for artificial intelligence \cite{Fjeld2020-oq,cole2024ai,Maslej2025AIIndex}, shaped by differing cultural values, governance models, and economic priorities. 

These frameworks fall broadly into binding laws and non-binding guidelines, both of which need to be considered for a comprehensive global understanding of AI governance. Binding laws establish enforceable obligations and penalties for non-compliance, typically categorizing AI systems by risk level and providing centralized oversight. Non-binding guidelines, in contrast, offer guidance without formal enforcement and are often paired with sectoral binding regulation, enabling agility, industry control, and iterative adaptation to rapidly evolving technologies. 

This paper analyzes regulatory approaches in China, North America, and Europe, highlighting both the diversity of governance philosophies and the impact of production-based factors on global AI research, assurance practices, and policy diffusion \cite{reuel2024positionpapertechnicalresearch,policydiffusion,siegmann2022brusselseffectartificialintelligence,Cheng03092023}. While some of these nations also have additional regulatory frameworks that impact AI assurance (sector-specific laws, cybersecurity and data governance initiatives, and industry standards) \cite{Maslej2025AIIndex}, this work focuses on comprehensive, general-purpose frameworks to provide a high-level comparative perspective.

\subsubsection{\textbf{Binding Laws}}

\begin{itemize}
    \item \textbf{European Union (EU) – AI Act:}
    
    Categorizes AI systems into four tiers: unacceptable, high, limited, and minimal risk. The Act adopts a precautionary, rights-based approach with centralized enforcement by EU authorities \cite{eu_ai_act}.

    \item \textbf{United States – Executive Order on Safe, Secure, and Trustworthy AI (2023):}

    Mandates evaluations of dual-use foundation models, protections for biometric data, and interagency coordination. It reflects the U.S. strategy of executive-led regulation in the absence of comprehensive federal legislation \cite{executive_order_14110}.

    \item \textbf{Canada – Artificial Intelligence and Data Act (AIDA):}

    Proposes enforceable obligations for “high-impact” systems, such as those used in employment or credit decisions. It seeks to balance transparency, accountability, and innovation, positioning itself between the EU’s stricter and the U.S.’s more flexible models \cite{canada_aida_2022}.

\end{itemize}

\subsubsection{\textbf{Non-Binding Guidelines}}

\begin{itemize}
    \item \textbf{United States – NIST AI Risk Management Framework (RMF):}
    Promotes trustworthy AI through transparency, accountability, and risk mitigation. Widely adopted by both federal agencies and private companies, it reflects the U.S. preference for sectoral self-regulation and innovation-driven governance \cite{NIST}.

    \item \textbf{United Kingdom – AI Regulation White Paper:} 
    Proposes a sector-specific model where existing regulators oversee AI use within their respective domains (e.g., healthcare, finance). It emphasizes principles such as transparency and favors agility and innovation over centralized mandates \cite{uk_ai_white_paper_2023}.

    \item \textbf{OECD – Principles on AI:}
    Provides high-level guidance on robustness, transparency, accountability, and fairness, endorsed by over 40 countries, serving as a reference point for national policies \cite{oecd_ai_principles_2024}.

    \item \textbf{China – AI Safety Governance Framework (2024):}
    Released by the National Information Security Standardization Technical Committee (TC260) that outlines principles for safe, responsible, and ethical AI development. It provides guidance for businesses, developers, and regulators on risk management, data governance, and model transparency \cite{china_framework}.
\end{itemize}

This work focuses on the EU AI Act and NIST RMF, which represent two of the most influential and contrasting approaches to AI governance: the former a binding, precautionary, rights-based model with centralized enforcement, while the latter illustrates a non-binding, voluntary approach emphasizing flexibility, sectoral self-regulation, and risk management guidance.

\section{Assurance Methodology}
\label{subsec:audit_methods}
AI governance is built upon methods for auditing, evaluation and analysis, largely shaped by the nature of existing AI systems and their underlying hardware. However, these methods are historically optimized for static, clock-driven computation – a poor fit for dynamic, spike-driven architectures that need to evolve at runtime. Yet, the features that make NeuroAI and neuromorphic systems intelligent and energy-efficient also result in difficulty utilizing these traditional governance methods. This section presents an overview of such methods and presents emerging limitations and challenges related to NeuroAI in neuromorphic systems.

\subsection{Traditional Approaches}
Work on AI auditing has sought to formalize systematic, replicable methods to assess model capabilities, safety, and societal impact. The following section highlights a set of common auditing assumptions, emerging from global regulatory frameworks, largely shaped by the nature of existing AI systems and their underlying hardware.

\textbf{Compute-Based Thresholds}

Computational power measured in FLOPs or similar metrics is often treated as a gauge for capability and risk. The EU AI Act defines $10^{25}$ FLOPs as the compute threshold for “systemic risks” \cite{eu_ai_act_article_51}. In the U.S., export controls use FLOP thresholds to restrict the development and distribution of high-performance AI chips and systems \cite{omb2025aiframwork}. The October 2023 U.S. AI Executive Order similarly mandates reporting of models trained with over $10^{26}$ FLOPs \cite{congress2025aiorder}. These thresholds serve as a common reference point for auditing AI.

\textbf{Models and Weight Inspection}

Many regulatory frameworks assume that a model’s internal logic, such as weights, layers, or decision paths, can be inspected, documented, and retained. For instance, Article 13 of the EU AI Act requires that high-risk AI systems maintain technical documentation detailing system architecture, training methodologies, and internal logic to ensure traceability and accountability \cite{eu_ai_act_article_13}. The NIST AI Risk Management Framework identifies traceability and transparency of internal processes as pillars of trustworthy AI, recommending organizations document model development, including intermediate artifacts like weights or training checkpoints.

Explainability has also become a globally measured metric through various methods as a form of procedural transparency, an assurance that system behavior can be rationalized and reconstructed when necessary \cite{dovsilovic2018explainable,xu2019explainable}. Canada’s AIDA emphasizes risk assessments that include internal system explanations, reinforcing the need for model interpretability \cite{canada_aida_2022}. The "Measure" function of the AI RMF Playbook further calls for quantitative and qualitative evaluations of explainability and robustness, which presuppose access to internal model parameters and architecture for analysis and benchmarking. Overall, frameworks reflect an underlying assumption that AI systems are architecturally and operationally transparent and that weights are static and loggable.

\textbf{Dataset Documentation and Provenance for Bias Identification}

Several frameworks place strong emphasis on dataset transparency, traceability, and quality. Article 10 of the EU AI Act mandates that datasets used in high-risk systems be representative, free of errors, and appropriately documented, including records of data sources, collection methods, and preprocessing techniques \cite{eu_ai_act_article_10}. The NIST AI RMF follows this, calling for rigorous dataset lifecycle documentation from acquisition to labeling to deployment \cite{NIST}. 

Meanwhile, the UK’s AI White Paper assigns oversight to sectoral regulators, who must ensure that data quality and provenance standards are upheld throughout the system’s lifecycle \cite{uk_ai_white_paper_2023}. The OECD AI Principles and China AI Safety Governance Framework similarly advocate for accountability and transparency, implicitly requiring that training datasets be auditable and non-obscure \cite{oecd_ai_principles_2024}. All in all, these frameworks rely on the assumption that datasets can be compartmentalized and traced.

\subsection{NeuroAI and Neuromorphic Computing}
Recent progress in neuromorphic computing has created a
technical movement seeking to establish standardized
benchmarking procedures. This shift reflects a growing
recognition that embodied, event-driven, and continuously
adaptive systems cannot be adequately assessed using
traditional AI metrics such as FLOPs or static accuracy
\cite{bartolozzi2022embodied}. A current effort leading
this movement is NeuroBench, proposed in 2023 by a
consortium of academic and industrial laboratories
\cite{neurobench}. NeuroBench represents the first attempt to
unify the assessment of neuromorphic algorithms, models,
and devices under a common methodological structure that
interfaces with conventional machine learning evaluation.

Before NeuroBench, benchmarking in neuromorphic
research was fragmented and device-specific. Performance
claims for hardware such as Intel’s Loihi
\cite{davies2018loihi}, IBM’s TrueNorth and NorthPole
\cite{modha2023science,appuswamy_breakthrough_2024}, or
large-scale mixed-signal systems like SpiNNaker
\cite{furber2014spinnaker} and BrainScaleS
\cite{schemmel2017brainscales} were reported in disparate
units, such as synaptic operations per second, joules per inference,
or spikes per neuron, making systematic comparison
impossible. Moreover, evaluation often conflated algorithmic
and hardware contributions, with each group selecting
proprietary tasks and metrics that favored its own
architecture. This heterogeneity mirrored the early years of
deep learning before ImageNet, impeding reproducibility,
comparability, and ultimately policy-relevant auditing.

NeuroBench addresses this gap by defining a
device-agnostic yet biologically meaningful evaluation
framework. Its stated goals are threefold:
\begin{enumerate}
    \item To provide a standardized task suite covering both
    conventional and neuromorphic workloads;
    \item To define cross-platform metrics capturing not only
    accuracy but also energy efficiency, latency, and
    event-driven performance;
    \item To create an open reference implementation enabling
    transparent reporting across research groups and hardware
    vendors.
\end{enumerate}
By explicitly linking algorithmic performance to hardware
efficiency, NeuroBench reframes benchmarking from a
purely computational exercise to a systems-level audit of
neuromorphic intelligence \cite{neurobench,bartolozzi2022embodied}. 

The NeuroBench framework organizes evaluation along
three hierarchical layers: task, model, and
platform, each accompanied by reproducible reference
code and measurement protocols:
\begin{itemize}
    \item \textbf{Task Layer:} Tasks span a spectrum from
    conventional static classification (e.g., MNIST, CIFAR-10)
    to event-based sensor processing (DVS Gesture,
    N-Caltech101), temporal prediction, and control tasks
    designed for continuous learning. This design acknowledges
    that neuromorphic systems excel on workloads characterized
    by temporal sparsity and event-driven dynamics rather than
    large-batch dense computation \cite{razzaq2025multimodal,bartolozzi2022embodied}.
    \item \textbf{Model Layer:} The framework supports both spiking
    neural networks (SNNs) and converted artificial neural
    networks (ANNs). This dual inclusion enables fair comparison
    between biologically grounded learning rules (e.g.,
    spike-timing-dependent plasticity) and ANN-to-SNN
    conversions optimized for hardware execution
    \cite{iqbal2025biologically,razzaq2025multimodal}.
    \item \textbf{Platform Layer:} Performance is evaluated across
    digital and mixed-signal neuromorphic chips, GPUs, and
    conventional CPUs. Each platform reports standardized
    telemetry, such as inference latency, power draw, total
    energy per sample, and resource utilization, facilitating
    device-independent benchmarking and longitudinal tracking
    of progress. 
\end{itemize}

A centralized results schema ensures that all submissions
include the same metrics and metadata, enabling reproducible,
cross-device comparison and transparent assessment of
neuromorphic progress. Emerging studies now extend this
framework to embodied and multimodal contexts, integrating
domain-generalized event–frame fusion and biologically
realistic cortical primitives within benchmarking pipelines
\cite{razzaq2025multimodal,iqbal2025biologically,bartolozzi2022embodied}. Collectively, these developments
illustrate the maturation of neuromorphic benchmarking from
isolated experiments into a coherent, scientifically grounded,
and policy-relevant methodology for the broader field of
NeuroAI.

\section{Governance Gaps}
\label{sec:govgaps}
Compute accounting has become the dominant proxy for capability and risk. The EU AI Act defines a threshold of approximately $10^{25}$ FLOPs for “systemic-risk” models. The 2023 U.S. Executive Order mandates reporting for training runs exceeding $10^{26}$ FLOPs, and export-control regimes employ similar limits \cite{omb2025aiframwork,executive_order_14110}. This practice presumes that computational throughput correlates with model capability.

Neuromorphic processors do not follow this premise. Their computation is event-driven and sparse, measured in spikes per second rather than synchronous floating-point operations.  Energy expenditure is proportional to the number of meaningful events, not to clock cycles. Reconstructing a heatmap of “what the network attended to” would require correlating spike trains across millions of neurons and microsecond-scale timestamps, producing results that are dynamic, non-stationary, and unintelligible in human time. In effect, the classical explainability stack relies on clocked continuity and feature independence, while neuromorphic systems are asynchronous, discrete, and path-dependent. Hence, auditability in such systems may require new instruments and visualization methods that capture temporal dependencies and emergent oscillatory patterns rather than static weight maps. A new interpretability regime must therefore emerge. Rather than feature attribution, auditability in NeuroAI demands dynamical-systems analysis: characterization of attractor landscapes, oscillatory coupling, spike synchrony, and stability margins. 

Typically, note that AI evaluation relies on a boundary between a system and its observers, treating models as black boxes. Auditors submit recorded inputs, observe outputs, and infer behaviors. A recent study showed that such black-box audits offer only a superficial understanding of system safety and can miss failures that appear only in internal dynamics . White-box audits, which inspect weights and activations, enable deeper interpretability but require access to stable, inspectable parameters. Outside-the-box audits go a step further, requiring documentation of training data, parameters, and development processes \cite{Casper_2024}.

Neuromorphic systems resist all three modes of auditing. With their state distributed across millions of plastic weights that constantly evolve with their environments, reproducing experiments becomes highly nuanced, as microscopic weight differences can alter entire network dynamics. Benchmarking model accuracy reveals little about safety, reliability, and interpretability. This makes neuromorphic systems harder to freeze, replicate, or inspect in accordance with the regulatory mechanisms described previously.

This section examines the principal auditing paradigms that shape existing regulation and explains why their methodological assumptions fail when applied to neuromorphic systems.

\begin{table}[ht]
  \centering
  \footnotesize
  \renewcommand{\arraystretch}{1.5} 
  \setlength{\tabcolsep}{4pt}
  
  \begin{tabular}{|>{\RaggedRight\arraybackslash}p{1.6cm}|
                  >{\RaggedRight\arraybackslash}p{1.9cm}|
                  >{\RaggedRight\arraybackslash}p{1.9cm}|
                  >{\RaggedRight\arraybackslash}p{1.8cm}|}
    \hline
    \textbf{Regulatory Lever} & \textbf{Conventional Assumption} & \textbf{Neuromorphic Reality} & \textbf{Governance Gap} \\
    \hline
    Compute thresholds & More FLOPs $\Rightarrow$ more capability / risk. & Event-driven spikes; tasks use minimal FLOPs. & Compute-based thresholds are inapplicable. \\
    \hline
    Inspectable weights & Static weights can be shared \& audited. & On-chip plasticity and local weight changes. & Audit and certification become obsolete. \\
    \hline
    Data governance & Training is separate; datasets are declared. & Continuous learning from ambient streams. & Consent and bias tracking become unclear. \\
    \hline
  \end{tabular}
  \caption{\textit{Neuromorphic hardware slips through AI regulations.}}
\end{table}

\subsection{Limits of FLOP-Centric Regulation}
Most governance frameworks currently treat computational power, typically measured in FLOPs (floating-point operations per second), as a proxy for model capability and risk \cite{gupta2024datacentricaigovernanceaddressing}. This metric has already been criticized for oversimplifying the relationship between compute and capability \cite{gupta2024datacentricaigovernanceaddressing,hooker2024limitationscomputethresholdsgovernance,watson2025beyond}. The problem becomes even more pronounced with neuromorphic systems, which differ fundamentally from conventional GPUs. However, this governance limitation further breaks down when applied to neuromorphic systems, which are fundamentally different in architecture and operation. Unlike conventional GPUs that rely on synchronous, clock-driven matrix multiplications, neuromorphic systems employ asynchronous, event-driven computation where spikes replace continuous activations and computation occurs only when triggered by relevant input.

As a result, FLOPs fail to meaningfully capture either the computational efficiency or the functional potential of these systems \cite{appuswamy_breakthrough_2024}. This mismatch has significant implications for regulation: policies based on FLOP thresholds may overlook powerful systems that operate outside this metric entirely. Despite using far less power, NorthPole achieves substantially higher energy efficiency and lower latency vs. performance metrics that FLOPs alone do not account for.

Following Appuswamy et al. (2024), the benchmark used a 3-billion-parameter transformer-decoder LLM derived from IBM’s Granite-8B-Code-Base model, consistent with architectures such as Llama 3 8B and Mistral 7B. The model was quantized to INT4 weights and activations (w4a4) for NorthPole, while the GPU baselines (L4, L40S, A100, H100) used FP16 weights and activations (w4a16) via GPTQ quantization and the vLLM 0.5.4 runtime with Marlin kernels. The NorthPole appliance comprised 16 PCIe Gen3~$\times$~8 cards in a 2U server, each with on-chip 192~MB SRAM and 13~TB/s memory bandwidth, executing inference with mini-batch = 28, micro-batches = 14a, and sequence length = 2048 tokens (1024 prompt + 1024 generated) using greedy decoding. NorthPole achieved 28,356 tokens/s system throughput and 1~ms/token latency while consuming 42~W per card (672~W total). When compared to the NVIDIA H100 (4~nm, 700~W) under its lowest-latency configuration, NorthPole delivered $\sim$72.7$\times$ higher energy efficiency (tokens/s/W) and $\sim$2.5$\times$ lower latency. FLOP-equivalent measures were not directly reported, as event-driven architectures lack a one-to-one mapping; approximate operation density from synaptic events suggests a reduction of about $10^2\times$ relative to H100 for comparable accuracy. These data provide empirical grounding for our claim that neuromorphic and event-driven efficiency gains are invisible to FLOP-based regulatory metrics yet verifiable through energy-per-token and latency-per-token benchmarks \cite{appuswamy_breakthrough_2024}.

\subsection{The End of Inspectable Weights and Systems}
Neuromorphic systems also introduce fundamental challenges to model transparency and auditability. Traditional AI models can often be examined through static, exportable weight matrices. In contrast, neuromorphic systems rely on on-chip plasticity and hardware-level learning that continuously adapts during inference \cite{bossard_analog_2024,davies2018loihi}. These updates occur in analog or low-level event-driven digital forms, making them inherently difficult to record, freeze, or reproduce \cite{davies2018loihi}.

\begin{figure}
    \centering
    \includegraphics[width=1\linewidth]{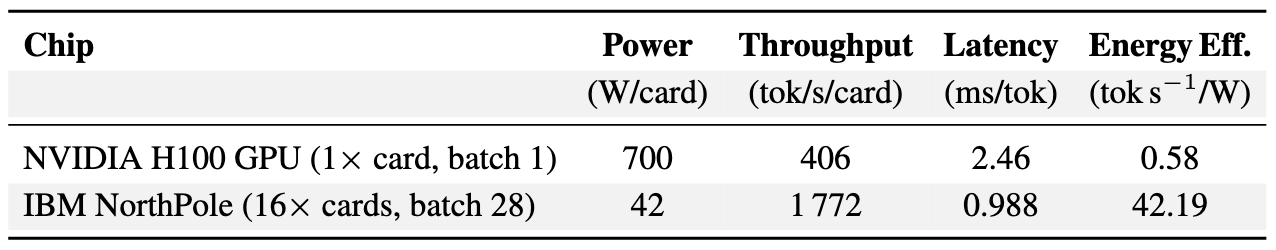}
    \caption{\textit{NorthPole performs at} \textbf{$\sim$70× higher energy efficiency} and \textbf{$\sim$2.5× lower latency} \textit{than H100 while using a non-FLOP metric, illustrating why neuromorphic computing can elude compute-based thresholds.}}
    \label{fig:placeholder}
\end{figure}

This dynamic behavior undermines the core assumptions in many regulatory frameworks, such as the EU AI Act and China AI Safety Governance Framework, which require that models be inspectable, documentable, and replicable. For instance, the NIST AI RMF’s “Govern” and “Measure” functions rely on audit trails and evaluation procedures that assume access to model internals, including weights and training checkpoints \cite{NIST}. In neuromorphic systems, such internal states may not be externally accessible or even meaningfully interpretable.

\subsection{Data in the Dark}
Neuromorphic architectures challenge long-standing assumptions in AI regulation by collapsing the boundary between training and inference. These systems, inspired by the mammalian brain, rely on continuous learning, updating in real time based on new inputs \cite{appuswamy_breakthrough_2024}. These systems may incorporate unlabeled or ambient data from their environment, complicating efforts to ensure informed consent, traceability, and fairness. The opacity of neuromorphic learning processes raises critical concerns about the introduction and amplification of bias, especially when such updates are invisible even to the developers who built the system.

Tracing the internal logic of modern AI systems is already difficult, and neuromorphic systems only magnify this problem \cite{christensen2022roadmap}. While some scholars argue that algorithmic bias can be measured and corrected more easily than human bias \cite{cdei2020bias}, this technical optimism does not resolve legal mandates for documentation and oversight. The EU AI Act, for example, mandates detailed documentation of data representativeness, labeling, and quality \cite{eu_ai_act}; similarly, the NIST AI Risk Management Framework stresses data traceability and informed data governance under its “Measure” and “Govern” functions \cite{NIST}.

NeuroAI under neuromorphic systems upends these requirements. Their unbounded learning cycles, lack of fixed training phases, and unpredictable interactions with real-world data render dataset compartmentalization and auditing virtually impossible. This collapses the regulatory distinction between model development and deployment, breaking the logic and assumptions underpinning most existing data governance tools.

\section{Discussion}
\label{sec:discussion}
\subsection{Limitations}
This analysis is limited by its focus on a subset of AI governance frameworks, with less attention to cultural and regional variations. Additionally, while broad regulatory approaches are considered, there is limited exploration of sector-specific assurance, such as healthcare laws in the U.S., where assurance methods and compliance requirements may differ \cite{Maslej2025AIIndex}. 

\subsection{Governance Takeaways}
The governance gaps introduced by neuromorphic computing are not just theoretical; they reveal a fundamental misalignment between emerging AI architectures and current regulatory assumptions. Today’s AI governance models rely heavily on characteristics of traditional von Neumann systems: performance metrics like FLOPs, exportable model weights and traceable datasets. However, NeuroAI and neuromorphic systems challenge each of these assumptions. These systems learn continuously, adapt locally, and operate through distributed, hardware-integrated memory, making them difficult to audit, export, or benchmark.

\subsection{High-Stakes Implications}
The urgency of addressing these governance challenges is growing as neuromorphic computing moves from laboratory settings into real-world applications. Neuromorphic systems are especially well-suited for edge computing scenarios that demand energy efficiency, low latency, and autonomous adaptation—characteristics critical in healthcare devices and autonomous vehicles \cite{DeWolf2020-tb}. These sectors not only heighten the stakes for governance but also complicate compliance, since such devices often operate in data-sensitive or safety-critical environments with limited connectivity or human oversight. As these architectures begin powering safety-critical applications, the governance conversation must move from hypothetical to urgent.
This raises a variety of regulatory implications:
\begin{itemize}
    \item \textbf{Safety and audit risks:} In domains like autonomous driving or medical diagnostics, continual on-device learning without a clear audit trail can result in behavior that is neither reproducible nor explainable, which can pose serious liability and accountability issues.
    \item \textbf{Global regulatory divergence:} AI risk tiering schemes, and consequently, export controls \cite{bis2025framework}, based on compute thresholds may entirely miss neuromorphic chips. This could result in uneven international regulation.
\end{itemize}

\subsection{Market Effects}
Neuromorphic chips promise steep gains in energy efficiency and real-time performance by co-locating memory and compute and by using event-driven spikes, making them attractive for modern computing. Yet their supply chain is still juvenile: fabrication runs are small, specialized mixed-signal design processes raise upfront engineering costs, and researchers continue to report obstacles in scalability and integration with conventional tools \cite{Elfighi2025-jw}. Regulatory requirements that mandate local production or certification can further amplify these challenges, as the associated costs must be incurred in each jurisdiction. The resulting effect is narrower vendor options, higher unit prices, and slower deployment of the low-power hardware that policymakers aim to promote \cite{Hancke2022-tc}.

\section{Future Work}
\label{sec:futurework}
Future research should pursue a co-development approach, aligning governance frameworks with emerging technical research. This shift will require collaboration between technical researchers, hardware designers, policy stakeholders, and regulators. Bridging governance and innovation depends on treating regulatory assurance as part of the design process, not an afterthought. 

We have a collection of recommendations for researchers in this space:
\begin{itemize}
    \item \textbf{Build Interdisciplinary Collaborations:} Governance should not be designed in isolation from the systems it aims to regulate. Researchers must engage not only with policy makers and ethicists, but also with the domain practitioners (such as healthcare workers) who will ultimately use or be affected by these technologies. 
    \item \textbf{Open Assurance from the Onset:} Researchers should embed transparency and assurance mechanisms directly into the design and research process. This involves developing shared frameworks and group platforms, similar to initiatives like EvalEval, which cultivate community-driven benchmarking and interpretability research through an open online slack group. By coordinating evaluation practices early and across a large community, we can collectively shape assurance that evolves alongside cutting-edge research.
    \item \textbf{Foster Transparency:} Researchers should communicate not just what their systems achieve, but how they behave under uncertainty and adaptation. Making development processes, data provenance, and architectural decisions publicly accessible enables policymakers, journalists, and affected communities to engage meaningfully in oversight. 
\end{itemize}

\subsection{Rethinking Metrics}
Current regulatory audit methods (FLOP-based thresholds, weight disclosures, and dataset audits) assume architectural characteristics that neuromorphic systems fundamentally disrupt. If regulatory frameworks are to remain effective, they must evolve alongside non-Von Neumann architectures. Rather than retrofitting outdated metrics, we must rethink how we evaluate performance, risk, and accountability.

To ensure we do not repeat the same roadblocks faced by current AI governance-where regulation lags behind technical capabilities, future metrics must be designed alongside the architectures they aim to govern \cite{DBLP:journals/corr/abs-1802-07228,Floridi2018-do}. This means bridging the persistent gap between regulatory expectations and technical research.

Two directions offer promising starting points:
\begin{itemize}
    \item \textbf{Energy per Inference:} Rather than counting FLOPs, regulators can require every edge-class accelerator to publish the \emph{integrated energy} (joules) consumed to produce one output on a standard benchmark \cite{neurobench}. Recent neuromorphic results illustrate the spread: Intel Loihi 2 consumes $\sim\!0.0025$\,mJ per MNIST inference, versus $\sim\!10$–$30$\,mJ on GPUs of similar accuracy \cite{Renner2024-ty}. 

    \item \textbf{Weight-Change Ratio:} Many neuromorphic chips can raise a one-bit ``flag'' whenever a learning rule updates a synapse. Sampling those flags every $N$ inferences yields the runtime statistic 
\[
\mbox{WTR} = \frac{\Delta W_{\mbox{flags}}}{W_{\mbox{total}}},
\] 
where $\Delta W_{\mbox{flags}}$ is the number of weights modified in the time window and $W_{\mbox{total}}$ is the total number of synapses. A low, steady WTR signals a mostly frozen model; a sudden spike reveals on-device adaptation, concept drift, or possible tampering. Publishing WTR in a log allows for throttling or sandboxing of any model whose plasticity exceeds a predefined threshold, without ever exporting raw weights \cite{neurobench}.

\end{itemize}

\subsection{Governance Recommendations}
The case of neuromorphic computing illustrates the dangers of regulation that lags behind technical innovation. If frameworks continue to rely on assumptions rooted in outdated architectures, they risk missing an entire class of intelligent systems, which leaves critical use cases unregulated, unaudited, and potentially unsafe.

To keep governance meaningful, researchers and policymakers alike must co-evolve their frameworks with the systems they aim to regulate.

In tandem with building efforts to foster interdisciplinary research communities, several potential governance directions emerge for further exploration:

\begin{itemize}    
    \item \textbf{Architecture-Aware Differentiation:} Governance should be sensitive not just to the functionality of a model, but also to the \textit{underlying architecture} on which it operates. This implies developing technical taxonomies for categorizing architectures by their governability (e.g., whether weights are static, learn continuously, or operate through analog state changes) and tailoring risk assessments accordingly.
    \item \textbf{Embedded Governance Primitives:} Future systems should include hardware-level governance affordances by design \cite{petrie2}. These could include secure enclaves for audit logging, federated traceability mechanisms, or hardwired energy ceilings. Such primitives would enable governance at runtime and reduce reliance on post hoc interpretability tools, especially in decentralized or edge deployments.
    \item \textbf{Cross-Border Regulatory Coordination:} As neuromorphic systems mature, their global development and deployment will outpace isolated national frameworks. Transnational coordination through OECD bodies, standard regulation, and bilateral research exchanges will be critical to building a cohesive governance approach that remains effective across computational, legal, and cultural contexts. Cross-border regulatory coordination must not only ensure equitable resource distribution but also deliberately include regions and actors historically excluded from global standard-setting \cite{Veale2023-ny}. 
\end{itemize}

Ultimately, future governance will need to shift from passive auditing to proactive co-design, embedding compliance into the core of system architecture. This shift demands not only broader regulatory scope but also deeper technical literacy among policymakers and governance bodies. At the same time, researchers, through their autonomy and proximity to innovation, hold a unique responsibility to critically interrogate emerging technologies and translate technical insight into actionable governance frameworks. By fostering continuous dialogue between technical and regulatory domains, we can ensure that the next generation of intelligent systems is developed responsibly, transparently, and in alignment with societal values.

\bibliography{references}
\bibliographystyle{aaai}

\end{document}